\documentclass[aps,prl,reprint,superscriptaddress,longbibliography,nofootinbib]{revtex4-2}

\newif\ifprintsupplement
\printsupplementtrue

\usepackage{graphicx}
\graphicspath{{./}{figures/}}
\usepackage{physics}
\usepackage{custom_acronym}
\usepackage{dsfont}
\usepackage{amsfonts}
\usepackage[colorlinks=true, linkcolor=blue, citecolor=blue, urlcolor=blue]{hyperref}
\usepackage{orcidlink}

\ifprintsupplement
  \newcommand{\citesupp}{}
\else
  \newcommand{\citesupp}{~\cite{supplemental_material}}
\fi

\let\keeptoday\today

\def\today{\keeptoday}

\begin{document}

\title{Bayesian Calibration of Gravitational-Wave Detectors Using Null Streams Without Waveform Assumptions}

\author{Isaac~C.~F.~Wong\,\orcidlink{0000-0003-2166-0027}}
\email{chunfung.wong@kuleuven.be}
\affiliation{KU Leuven, Department of Electrical Engineering (ESAT), STADIUS Center for Dynamical Systems, Signal Processing and Data Analytics, Kasteelpark Arenberg 10, 3001 Leuven, Belgium}
\affiliation{Leuven Gravity Institute, KU Leuven, Celestijnenlaan 200D box 2415, 3001 Leuven, Belgium}
\author{Francesco~Cireddu\,\orcidlink{0009-0002-7074-4278}}
\affiliation{Leuven Gravity Institute, KU Leuven, Celestijnenlaan 200D box 2415, 3001 Leuven, Belgium}
\affiliation{Department of Physics and Astronomy, Laboratory for Semiconductor Physics, KU Leuven, B-3001 Leuven, Belgium}
\author{Milan~Wils\,\orcidlink{0000-0002-1544-7193}}
\affiliation{Leuven Gravity Institute, KU Leuven, Celestijnenlaan 200D box 2415, 3001 Leuven, Belgium}
\affiliation{Department of Physics and Astronomy, Laboratory for Semiconductor Physics, KU Leuven, B-3001 Leuven, Belgium}
\author{Tom~Colemont\,\orcidlink{0009-0003-4408-5894}}
\affiliation{KU Leuven, Department of Electrical Engineering (ESAT), STADIUS Center for Dynamical Systems, Signal Processing and Data Analytics, Kasteelpark Arenberg 10, 3001 Leuven, Belgium}
\affiliation{Leuven Gravity Institute, KU Leuven, Celestijnenlaan 200D box 2415, 3001 Leuven, Belgium}
\author{Harsh~Narola\,\orcidlink{0000-0001-9161-7919}}
\affiliation{Institute for Gravitational and Subatomic Physics (GRASP), Utrecht University, Princetonplein 1, 3584 CC Utrecht, The Netherlands}
\affiliation{Nikhef -- National Institute for Subatomic Physics, Science Park 105, 1098 XG Amsterdam, The Netherlands}
\author{Chris~Van~Den~Broeck\,\orcidlink{0000-0001-6800-4006}}
\affiliation{Institute for Gravitational and Subatomic Physics (GRASP), Utrecht University, Princetonplein 1, 3584 CC Utrecht, The Netherlands}
\affiliation{Nikhef -- National Institute for Subatomic Physics, Science Park 105, 1098 XG Amsterdam, The Netherlands}
\author{Tjonnie~G.~F.~Li\,\orcidlink{0000-0003-4297-7365}}
\affiliation{KU Leuven, Department of Electrical Engineering (ESAT), STADIUS Center for Dynamical Systems, Signal Processing and Data Analytics, Kasteelpark Arenberg 10, 3001 Leuven, Belgium}
\affiliation{Leuven Gravity Institute, KU Leuven, Celestijnenlaan 200D box 2415, 3001 Leuven, Belgium}
\affiliation{Department of Physics and Astronomy, Laboratory for Semiconductor Physics, KU Leuven, B-3001 Leuven, Belgium}

\date{\today}

\begin{abstract}
	We introduce a Bayesian null-stream method to constrain calibration errors in closed-geometry gravitational-wave (GW) detector networks.
	Unlike prior methods requiring electromagnetic counterparts or waveform models, this method uses sky-independent null streams to calibrate the detectors with any GW signals, independent of general relativity or waveform assumptions.
	We show a proof-of-concept study to demonstrate the feasibility of the method.
	We discuss prospects for next-generation detectors like Einstein Telescope, Cosmic Explorer, and LISA, where enhanced calibration accuracy will advance low-frequency GW science.
\end{abstract}

\maketitle

\textit{Introduction---} The detection of \cglspl{GW} has revolutionized astrophysics, enabling the observation of cosmic phenomena such as \cgls{BBH}, \cgls{NSBH}, and \cgls{BNS} mergers~\cite{1811.12907,2010.14527,2108.01045,2111.03606}.
Ground-based \cgls{GW} detectors, such as those operated by the \cgls{LVK} collaboration~\cite{1411.4547,1408.3978,2005.05574}, rely on precise measurements of spacetime distortions to extract astrophysical signals.
Central to this process is the accurate calibration of these detectors~\cite{1007.3973,1602.03845,1708.03023,2005.02531,2107.03294,2009.09305,2508.08423}, which ensures that observed data reliably reflect the true \cgls{GW} signals.
The \cgls{ET}~\cite{2010CQGra..27h4007P}, a next-generation detector with a triangular design, promises unprecedented sensitivity and the ability to construct a sky-independent null stream~\cite{1989PhRvD..40.3884G,0804.1036,0908.3665,2003.07375,2105.09485,2108.05108,2204.08533,2411.15506} --- a linear combination of detector outputs that cancels \cgls{GW} signals regardless of the sky position of the sources --- providing a novel opportunity for detector calibration.
Similarly, the \cgls{LISA}~\cite{2402.07571}, a space-based \cgls{GW} observatory, also employs a triangular configuration to form a null stream~\cite{2108.02738}, enabling complementary calibration techniques for low-frequency \cgls{GW} signals.
This work explores the use of the null stream, leveraging astrophysical signals to measure and correct calibration errors in \cgls{GW} detectors.

\cgls{GW} detectors, such as large-scale laser interferometers, do not directly measure the \cgls{GW} strain $h$.
Instead, the interferometer's output is a digital signal, $d_{\mathrm{err}}$, which represents the residual differential arm length, $\Delta L_{\mathrm{res}}$, after active feedback systems have attempted to correct for the displacement of the test masses.
The primary goal of calibration is to determine the detector's frequency-dependent response function, $R(f)$, which is the transfer function that accurately converts the digital output signal, $\tilde{d}_{\mathrm{err}}(f)$, to the \cgls{GW} strain in the frequency domain, $\tilde{d}(f) = R(f)\tilde{d}_{\mathrm{err}}(f)$.
This response function incorporates the complex optomechanical and electronic components of the detector.
Current methods rely on invasive, in-situ techniques like \cglspl{PCal}~\cite{1602.03845,1708.03023,2005.02531,2107.03294,2009.09305,2508.08423} and \cglspl{NCal}~\cite{gr-qc/0701134,1804.08249,1806.06572,2107.00141,2508.08423,2406.10028}, which inject known forces onto the test masses to measure the detector's response.
Alternative non-invasive methods, such as astrophysical calibration using known \cgls{GW} sources based on \cgls{EM} observations, have been proposed to continuously monitor the detector's response without interrupting observations~\cite{1511.02758,1902.08076}.
In practice, for LIGO, calibration errors are typically estimated on an hourly cadence~\cite{2005.02531}.
For Virgo, calibration errors are estimated for the entire observing run in O2~\cite{1807.03275}, while in O3, they are estimated separately for the O3a and O3b sub-runs~\cite{2107.03294}.

The impact of detector calibration errors on \cgls{PE} in \cgls{GW} astronomy has been extensively studied.
Early work, such as~\cite{1111.3044}, forecasted calibration error effects on Bayesian parameter estimation for binary inspiral systems in the advanced detector era, while~\cite{1712.09719} used the Cramér-Rao bound to establish systematic calibration requirements for \cgls{GW} detectors.
More recent studies provide insights into current and future challenges.
Ref.~\cite{2202.00823} demonstrated that calibration errors in operational ground-based detectors can degrade signal detection and parameter estimation accuracy.
For isotropic \cgls{GW} backgrounds, Ref.~\cite{2301.13531} showed that calibration errors distort detection and cosmological parameter inference.
In the context of future space-based detectors, Ref.~\cite{2204.13405} highlighted how calibration errors could limit \cgls{LISA}'s ability to probe fundamental physics.
Ref.~\cite{2204.03614} found that calibration errors bias Hubble constant measurements from \cgls{GW} sources, impacting cosmological studies.
Ref.~\cite{2506.15979} further noted that calibration accuracy affects black hole spectroscopy, critical for testing \cgls{GR} via ringdown signals.

Achieving high calibration accuracy is paramount for next-generation \cgls{GW} detectors to realize their potential for precision science, including cosmological parameter estimation and tests of fundamental physics.
As noted earlier, the triangular configuration of \cgls{ET} and \cgls{LISA} enables a sky-independent null stream, which cancels \cgls{GW} signals regardless of source location.
This null stream facilitates a novel, non-invasive astrophysical calibration approach that leverages \cgls{GW} signals themselves to measure and correct calibration errors, without relying on \cgls{EM} observations or assuming the validity of \cgls{GR}.
By complementing existing calibration techniques, this method enhances the precision of \cgls{GW} detector calibration, offering a robust tool to improve the accuracy of astrophysical measurements across next-generation observatories.

\textit{Sky-independent null stream---} Null stream is a linear combination of strain data from a network of \cgls{GW} detectors that cancel \cgls{GW} signals, regardless of their morphology, under ideal conditions~\cite{1989PhRvD..40.3884G}.
For general networks, null stream construction requires the source's sky position, as detector responses depend on \cgls{GW} direction.
However, for closed-geometry networks like triangles, null stream is sky-independent in the long-wavelength approximation, where the sum of detector tensors $\vb{D}_{j}$ vanishes: $\sum_{j=1}^{3} \vb{D}_{j} = \vb{0}$~\cite{2108.05108}, which holds at low frequencies.
For the \cgls{ET}, with 10-km arms, it is valid below $\sim 1$~kHz~\cite{2412.01693}.
For \cgls{LISA}, with $2.5\times 10^{6}$~km arms, it breaks above $\sim 1$~mHz~\cite{2303.15929}.
These thresholds are order-of-magnitude estimates, as systematic errors exceeding statistical uncertainties in null-stream analyses require further investigation.
Within the frequency range where the sky-independent null stream is accurate, it enables \cgls{GW} detector calibration using astrophysical signals without waveform assumptions.
Moreover, the null stream cancels arbitrary \cgls{GW} polarizations, including scalar, vector, and tensor modes from modified gravity theories, as shown in Eq.~(5) of~\cite{2108.05108}.
Unlike calibration methods relying on \cgls{EM} observations and assuming \cgls{GR}~\cite{1511.02758,1902.08076}, null-stream-based calibration is \cgls{GR}-independent, preserving the detector network's ability to test fundamental physics without bias.

\textit{Self-calibration---} Ref.~\cite{2009.10212} introduced self-calibration for \cgls{GW} detector networks, using a Sagnac-based null stream in a triangular configuration,
\begin{equation}
	\begin{aligned}
		\tilde{d}_{\mathrm{Sagnac}}(f) & = \frac{1}{\sqrt{3}}\sum_{i = 1}^{3}\tilde{d}_{i}(f)                                       \\
		                               & =\tilde{n}_{\mathrm{Sagnac}}(f) + \frac{1}{\sqrt{3}}\sum_{i=1}^{3}C_{i}(f)\tilde{s}_{i}(f)
	\end{aligned}
\end{equation}
where $\tilde{d}_{i}(f)$ is the strain data from the $i$th detector, $\tilde{n}_{\mathrm{Sagnac}}(f) = \frac{1}{\sqrt{3}}\sum_{i=1}^{3}\tilde{n}_{i}(f)$ is the Sagnac combination of the detector noise $\tilde{n}_{i}(f)$ from the individual detectors, $C_{i}(f)$ is the complex calibration error, and $\tilde{s}_{i}(f)$ is the \cgls{GW} signal in the $i$th detector.
A matched filtering analysis is then performed on this Sagnac combination to estimate the calibration errors.
This approach requires explicit waveform templates to estimate calibration errors, which inherently relies on accurate theoretical waveform predictions and assumes the validity of \cgls{GR}, limiting its model independence.

\textit{Null-stream calibration---} In this work, we propose a model-independent approach, modifying the null-stream construction to account for calibration errors without waveform assumptions.
Our approach leverages GW signals to infer calibration errors without assuming their morphology or \cgls{GR}, enabling unbiased tests of fundamental physics.

The observation model for a network of \cgls{GW} detectors is
\begin{equation}
	\tilde{\vb{d}}(f) = \vb{C}(f; \boldsymbol{\theta})\vb{F}\tilde{\vb{h}}(f) + \tilde{\vb{n}}(f)\,,
	\label{eq:observation_model}
\end{equation}
where $\tilde{\vb{d}}(f)$ is the frequency-domain strain data, $\vb{C}(f; \boldsymbol{\theta})$ a diagonal matrix of complex calibration errors for each detector parameterized by $\boldsymbol{\theta}$, $\vb{F}$ the antenna pattern matrix, $\tilde{\vb{h}}(f)$ the \cgls{GW} polarizations, and $\tilde{\vb{n}}(f)$ the detector noise with power spectral density matrix $\vb{S}_{n}(f)$~\cite{0908.3665,2003.07375,2105.09485,2312.14614}.
The impact of calibration errors on the detector noise is absorbed into $\tilde{\vb{n}}(f)$ and therefore $\vb{S}_{n}(f)$.
In this work, we only consider a network of three detectors.

Following~\cite{0908.3665,2003.07375,2105.09485}, we construct the null stream via a projection matrix:
\begin{equation}
	\begin{aligned}
		 & \vb{P}_{\mathrm{null}}(f; \boldsymbol{\theta}) \\
		 & =
		\vb{I}_{3} -
		\vb{W}(f; \boldsymbol{\theta})
		[\vb{W}^{\dagger}(f; \boldsymbol{\theta})\vb{W}(f; \boldsymbol{\theta})]^{-1}
		\vb{W}^{\dagger}(f; \boldsymbol{\theta})
		\label{eq:null_projector}
	\end{aligned}
\end{equation}
where $\vb{I}_{3}$ is a $3\times 3$ identity matrix, $\vb{W}(f; \boldsymbol{\theta}) = \sqrt{2\Delta f}\vb{S}_{n}^{-1/2}(f)\vb{C}(f;\boldsymbol{\theta})\vb{F}$, and $\Delta f= 1 / T$ is the frequency resolution for data duration $T$.
The null stream is given by $\tilde{\vb{z}}(f) = \vb{P}_{\mathrm{null}}(f; \boldsymbol{\theta})\tilde{\vb{d}}_{w}(f)$, where $\tilde{\vb{d}}_{w}(f) = \sqrt{2\Delta f}\vb{S}_{n}^{-1/2}(f)\tilde{\vb{d}}(f)$ is the whitened strain data.
With true calibration parameters ($\boldsymbol{\theta} = \boldsymbol{\theta}_{\mathrm{true}}$), it contains no \cgls{GW} signals and follows a standard complex normal distribution in a one-dimensional subspace.

To enhance \cgls{PE} and mitigate noise transients, we transform the null stream to the time-frequency domain via an orthogonal wavelet transform, yielding $\tilde{\vb{z}}_{\mathrm{TF}}(t, f; \boldsymbol{\theta})$~\cite{2012JPhCS.363a2032N}.
The log-likelihood is
\begin{equation}
	\log p(\vb{d} | \boldsymbol{\theta})
	= \sum_{j,k\in\mathcal{I}}-\frac{1}{2}\abs{\tilde{\vb{z}}_{\mathrm{TF}}(t_{j}, f_{k}; \boldsymbol{\theta})}^{2} + \mathrm{constant}
	\label{eq:log_likelihood}
\end{equation}
where $\mathcal{I}$ selects time-frequency regions containing \cgls{GW} signals.

This approach is equally applicable to triangular \cgls{ET} and \cgls{LISA}, leveraging low-frequency \cgls{GW} signals to enhance calibration precision, thereby broadening the framework's impact across next-generation \cgls{GW} observatories.

\textit{Results---}
To evaluate the efficacy of our Bayesian null-stream method for constraining \cgls{GW} detector calibration errors, we conduct simulations within a triangular \cgls{ET} network.
We analyze 15 data segments, each containing a \cgls{BBH} signal generated with the \texttt{IMPhenomXPHM} waveform model~\cite{2004.06503}, with the source parameters detailed in the Supplemental Material\citesupp.
The luminosity distance is adjusted to yield network \cglspl{SNR} uniformly spanning 20 to 300, typical for the triangular \cgls{ET}~\cite{2303.15923}.
This setup aims to test the degree of calibration error constraints across diverse signal strengths, as stronger signals are expected to yield tighter constraints due to a larger calibration-induced signal leakage to the null stream.
We model calibration errors $\vb{C}(f; \boldsymbol{\theta})$ using cubic splines for amplitude and phase per detector, defined as $C_{j}(f_{k}) = (1 + \delta A_{j}(f_{k}))e^{i\delta \phi_{j}(f_{k})}$ for the $j$th detector at frequency $f_{k}$ corresponding to the $k$th spline knot~\cite{farr14b}.
The error at each spline knot is drawn from Gaussian distributions informed by the calibration uncertainties of LIGO Livingston (median and $1\sigma$ values) over the one-hour interval beginning at an arbitrarily chosen GPS time 1256894100, during the third observing run~\cite{ligo_virgo_calibration_uncertainty}.
Given the relative stability of calibration uncertainties throughout the observing run, we adopt the uncertainties at this GPS time as representative.
To evaluate the method's effectiveness in constraining calibration errors, we selected the realization of calibration errors that maximizes the null stream \cgls{SNR}, corresponding to the most significant calibration-induced signal leakage. This choice, essential for a proof-of-concept study, ensures that improvements in calibration precision are evident, as small calibration errors with low null stream \cgls{SNR} would produce limited constraints that are difficult to discern.
Detector noise, assumed uncorrelated between detectors, is simulated using the ET-D sensitivity curve~\cite{1012.0908}.
We use the same calibration errors and noise realizations across all 15 simulations to isolate how calibration constraints vary solely with the strength of signal leakage into the null stream.

The method presented in this paper is implemented in our \texttt{nullcal} package~\cite{nullcal}, which defines the log-likelihood function in Eq.~\eqref{eq:log_likelihood}, and Bayesian inference is performed with \texttt{Bilby}~\cite{1811.02042, 2024zndo..14025463T} using the \texttt{dynesty} sampler~\cite{1904.02180,2024zndo..12537467K,2004AIPC..735..395S,10.1214/06-BA127,0809.3437}.
Calibration errors for the three \cgls{ET} detectors are modeled independently, each parameterized by a cubic spline with 10 knots uniformly spaced in $\log f$
from 8 Hz to 512 Hz, following ~\cite{farr14b}.
The calibration parameters $\boldsymbol{\theta}$ represent the amplitude and phase errors at these knots.
For the prior settings, we assign Gaussian priors to $\boldsymbol{\theta}$ at each knot, with means and standard deviations set to the median and 1$\sigma$ uncertainties from LIGO Livingston's calibration uncertainties at GPS time 1256894100 during the third observing run, identical to the distributions used to generate the mock data's calibration error realizations.
These priors reflect constraints from instrumental methods, such as \cgls{PCal}, enabling us to quantify the additional precision gained from \cgls{GW} signals, complementing rather than replacing instrumental calibration techniques.
Details of the mock data generation and sampler settings are provided in the Supplemental Material\citesupp.

\begin{figure}
	\includegraphics[width=\linewidth]{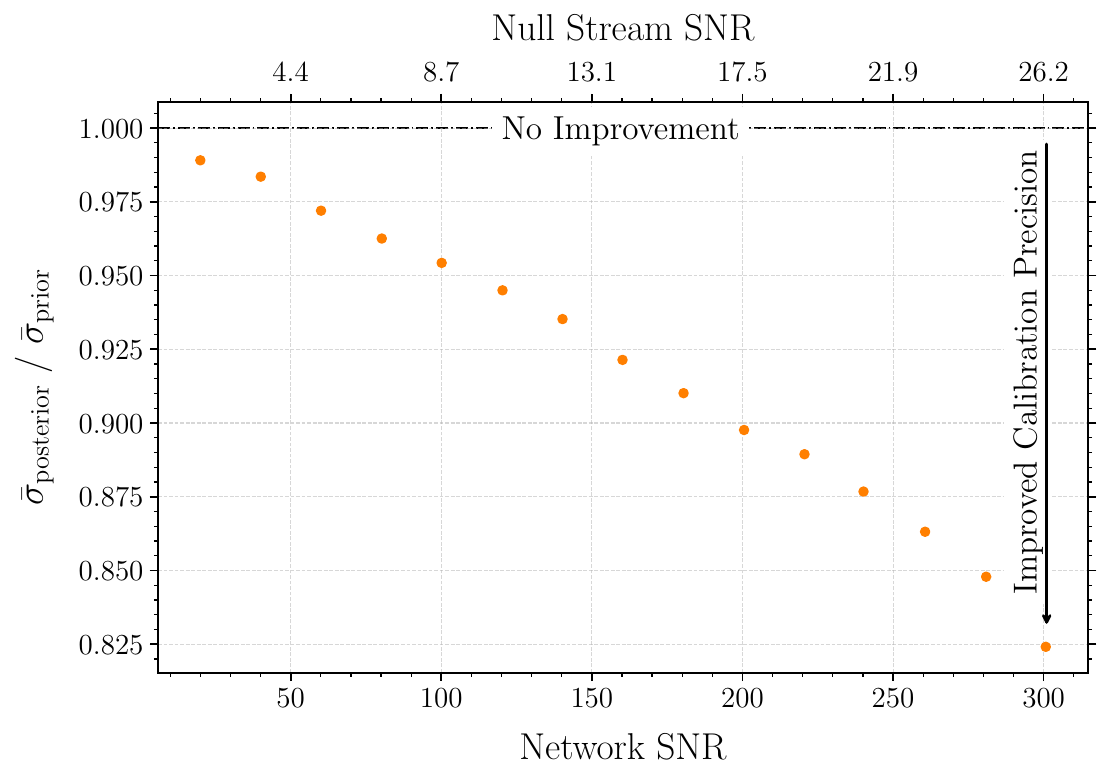}
	\caption{
		The plot shows the variation of the ratio of the posterior to prior generalized standard deviation of calibration parameters with increasing network \cgls{SNR} of the signal.
		The generalized standard deviation is defined as the determinant of the covariance matrix raised to $1 / (2N)$, where $N$ is the number of parameters.
		A ratio below $1$ indicates improved constraints, with $y = 1$ marking no improvement.
		As network \cgls{SNR} increases from 20 to 300, with the corresponding null stream \cgls{SNR} leakage shown at the top, the ratio decreases approximately linearly, demonstrating tighter constraints.
	}
	\label{fig:geometric_mean_volume_ratio_total}
\end{figure}

We quantify the calibration constraint by computing the geometric mean of the generalized standard deviation $\bar{\sigma}$~\cite{generalized_variance}, defined as the determinant of the covariance matrix of the distribution of the calibration parameters raised to $1 / (2N)$, where $N$ is the number of calibration parameters.
This metric, equivalent to the square root of the geometric mean of the covariance matrix eigenvalues, represents the typical scale of uncertainty per parameter dimension in the multidimensional parameter space.
Figure~\ref{fig:geometric_mean_volume_ratio_total} illustrates the performance of our method by plotting the ratio of $\bar{\sigma}$ of posterior samples to that of prior samples as a function of network \cgls{SNR}, ranging from 20 to 300.
The y-axis shows a linear decrease of this ratio with increasing \cgls{SNR}, reflecting progressively tightly calibration constraints as stronger signals increase the \cgls{SNR} of the calibration-induced signal leakage into the null stream, presented as the null stream \cgls{SNR} on the top axis, ranging from 1.75 to 26.3.
For the chosen set of calibration errors and \cgls{GW} signals, the results demonstrate up to 17.6\% improvement when \cgls{GW} signals are used in addition to traditional instrumental calibration techniques modeled as the prior in this study.

Degeneracy: The null stream method exhibits two key degeneracies affecting calibration error constraints.
First, a common mode degeneracy arises because the likelihood, incorporating the projection matrix $\vb{P}_{\mathrm{null}}(f;\boldsymbol{\theta})$ in Eq.~\eqref{eq:null_projector}, is invariant under a common scaling of the calibration errors $\vb{C}(f; \boldsymbol{\theta})$ across detectors, mimicking a modification of the source's luminosity distance~\cite{2009.10212}.
Consequently, relative calibration errors between detectors are constrained more tightly than absolute errors.
This method complements, rather than replaces, instrumental calibration techniques.
Instrumental measurements, encoded as priors, partially break this degeneracy, ensuring informative posteriors for absolute errors, though relative errors remain better constrained.
Second, when the \cgls{GW} signal is polarized i.e.\ $\tilde{h}_{\times}(f) = \varepsilon e^{i\psi}\tilde{h}_{+}(f)$, with ellipticity $\varepsilon$ and orientation angle $\psi$, two equivalent solutions for $\vb{C}(f; \boldsymbol{\theta})$ exist (up to a scaling factor), leading to broader posteriors.
Detailed analysis is provided in the Supplemental Material\citesupp.

\begin{figure}
	\includegraphics[width=\linewidth]{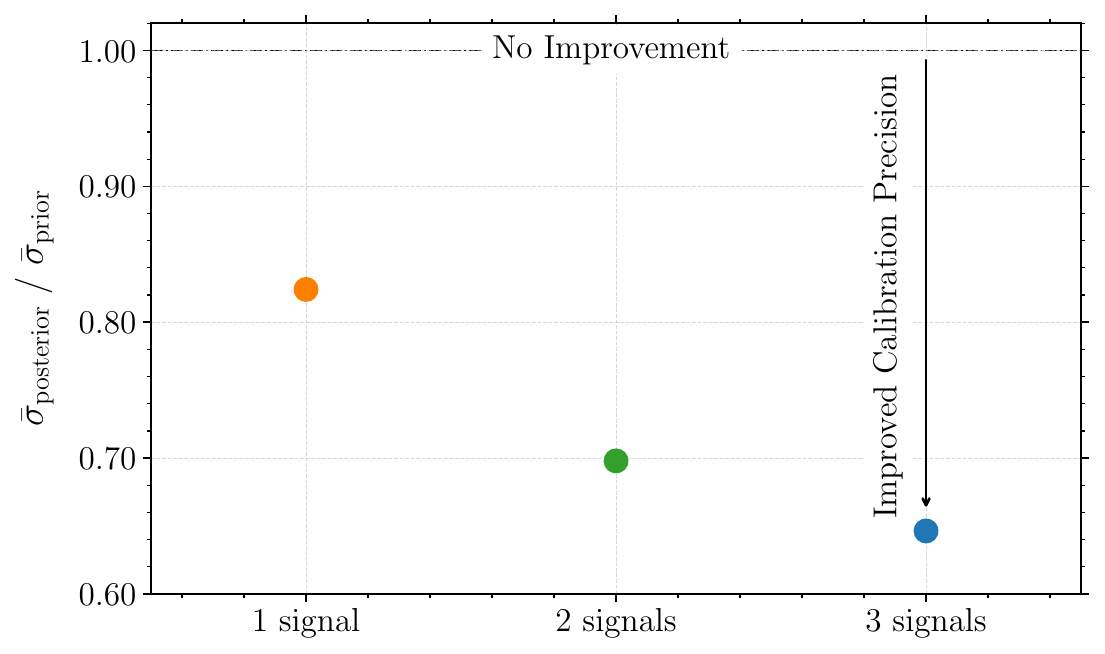}
	\caption{
		The plot shows the ratio of the posterior to prior generalized standard deviation of the calibration parameters as a function of the number of signals (1--3).
		The generalized standard deviation is defined as the determinant of the covariance matrix raised to $1 / (2N)$, where $N$ is the number of parameters.
		A ratio below $1$ indicates improved constraints, with $y = 1$ marking no improvement.
		The single-signal case has a null stream \cgls{SNR} of 26.3, with signals in the 2- and 3-signal cases scaled to maintain this null stream \cgls{SNR}.
		Increasing the number of signals from one to three tightens the posterior, breaking polarization degeneracies, consistent with the theoretical expectations.
	}
	\label{fig:geometric_mean_volume_ratio_total_comparison}
\end{figure}

The degeneracy is mitigated by higher-order harmonics in the \cgls{GW} emission or, more effectively for next-generation detectors like \cgls{ET} and \cgls{LISA}, by multiple \cgls{GW} signals overlapping in the frequency domain.
Figure~\ref{fig:geometric_mean_volume_ratio_total_comparison} shows the ratio of $\bar{\sigma}$ of posterior to prior samples for setups with one to three \cgls{GW} signals, each scaled to a total null stream \cgls{SNR} of 26.3, as detailed in the Supplemental Material\citesupp.
The ratio decreases noticeably with more signals: from 0.824 with a single signal, to 0.698 with two signals, and to 0.646 with three signals. The most substantial gain occurs when going from one to two signals due to greater degeneracy breaking for polarized signals. Beyond two signals, the improvement diminishes, leveling off as additional signals contribute less to resolving the degeneracy.

\begin{figure}
	\centering
	\includegraphics[width=\linewidth]{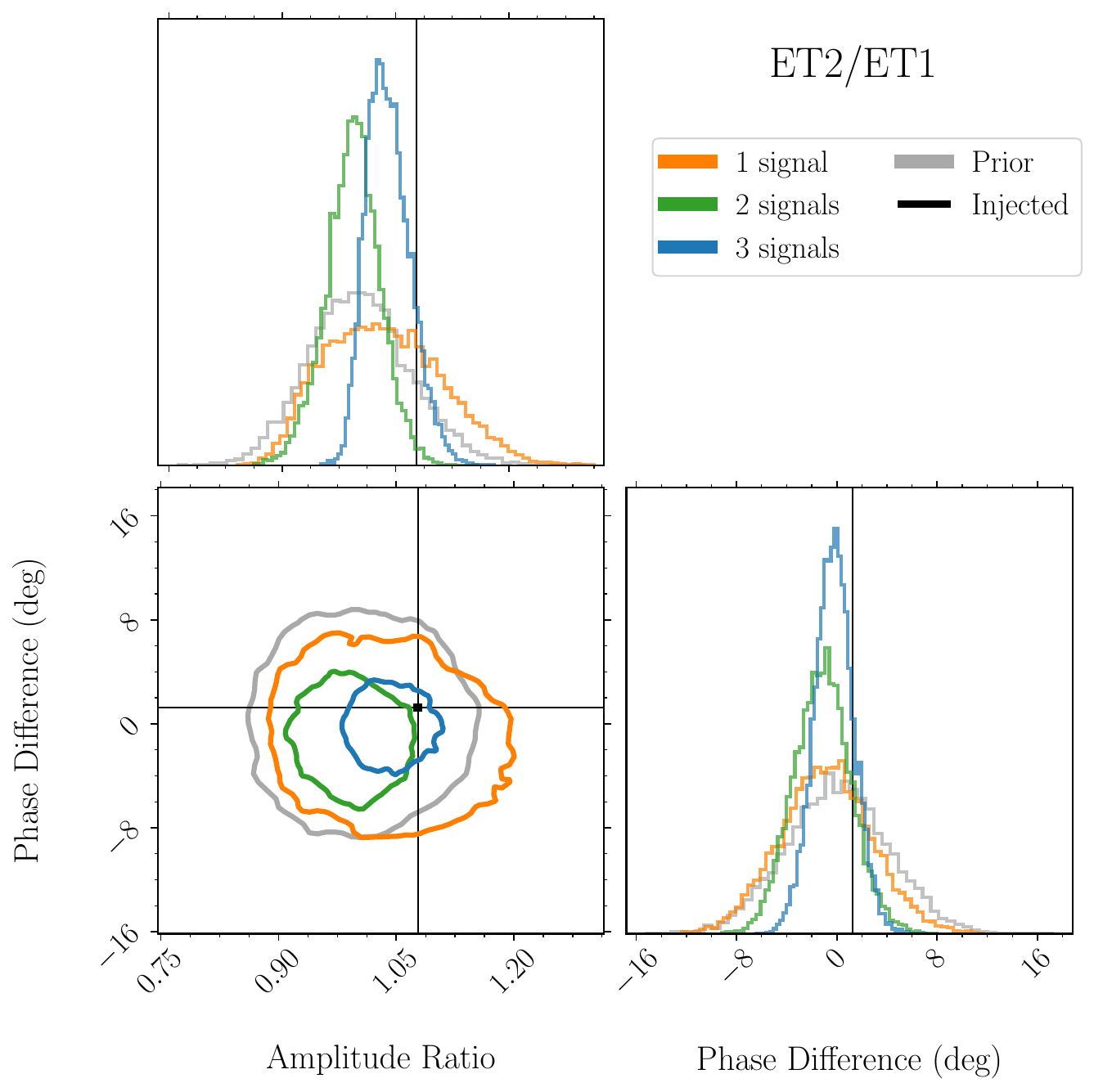}\\
	\includegraphics[width=\linewidth]{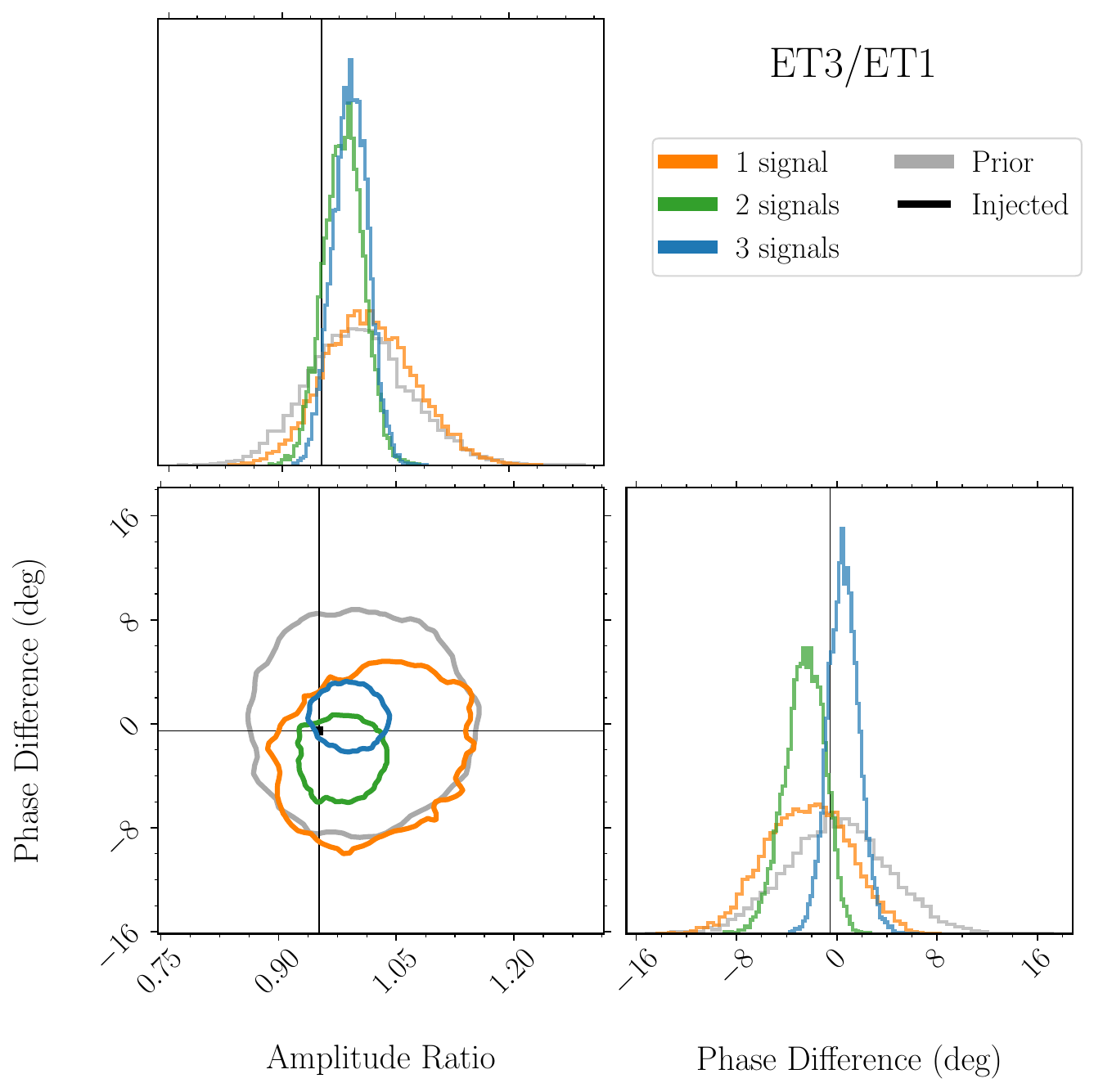}
	\caption{
		The plot shows the posterior distributions for the amplitude ratio $(1 + \delta A_{j}) / (1 + \delta A_{1})$ and phase difference $\delta \phi_{j} - \delta \phi_{1}$ of calibration errors for the $j$th \cgls{ET} detector compared to the first, at 32 Hz, where $\delta A_{j}$ and $\delta \phi_{j}$ are the amplitude and phase errors, respectively.
		The top panel compares the second detector to the first, and the bottom panel compares the third to the first.
		Contours represent 90\% credible regions.
		As the number of signals increases from one to three (null stream \cgls{SNR} fixed at 26.3), the posteriors tighten significantly, converging to the true values.
	}
	\label{fig:relative_calibration_error_marginal_comparison}
\end{figure}

In Fig.~\ref{fig:relative_calibration_error_marginal_comparison}, we present the posterior distributions of the relative calibration errors, defined as the amplitude error ratio $(1 + \delta A_j) / (1 + \delta A_1) $ and phase error difference $\delta \phi_j - \delta \phi_1$ with respect to the first \cgls{ET} detector, at 32 Hz for the \cgls{BBH} signals with a null stream \cgls{SNR} of 26.3.
This example illustrates how constraints on calibration errors improve with increasing numbers of signals, with full-frequency results (8--512 Hz) provided in the Supplemental Material\citesupp.

\textit{Conclusion and perspectives---} In conclusion, this work significantly advances the concept of self-calibration proposed by~\cite{2009.10212} for closed-geometry \cgls{GW} detector networks.
Building on this idea, we present a different strategy and develop the method into a model-independent framework by leveraging null-stream analysis, in contrast to the matched-filtering-based approach of~\cite{2009.10212}, which relies on specific waveform assumptions.
Unlike prior methods requiring \cgls{EM} counterparts and waveform assumptions, this approach leverages sky-independent null streams to utilize all \cgls{GW} signals, including astrophysical events and cosmological backgrounds, free from systematics due to waveform and \cgls{GR} assumptions.
We have conducted experiments to demonstrate that constraints on calibration errors tighten linearly with increasing network \cgls{SNR} (20--300) and multiple frequency-overlapping signals, breaking the degeneracy from polarizations, as detailed in the Supplemental Material\citesupp.
We also provide a practical implementation of the framework in the \texttt{nullcal} package~\cite{nullcal}.

This precision is critical for next-generation \cgls{GW} detectors like \cgls{ET} and \cgls{CE}, guiding optimal network geometry design to leverage the sky-independent null stream for precise calibration of the detectors, and for \cgls{LISA}, where the T channel being insensitive to \cgls{GW} signals below $\sim1\,\mathrm{mHz}$ enables the application of this method at low frequencies.
Enhanced calibration precision at low frequencies is critical for next-generation \cgls{GW} detectors, enabling robust scientific studies of low-frequency sources, such as precise \cgls{PE} for massive black hole binaries, early-warning alerts for binary neutron star mergers, and detection of primordial black holes, thus advancing \cgls{GW} astronomy~\cite{2503.12263}.

The proof-of-concept study demonstrates the feasibility of the method to constrain calibration uncertainties.
However, the corresponding \cgls{PE} improvements for source properties depend on calibration error frequencies and signal content, requiring careful experimental design to give robust conclusions.
Specifically, using the same data segment to constrain calibration uncertainties and perform \cgls{PE} double-counts the \cgls{GW} signal information, leading to overestimated precision or systematic errors.
Instead, off-source data must be used to measure calibration uncertainties, ensuring the uncertainty of \cgls{PE} on on-source signals is not underestimated.
A large-scale \cgls{MDC} with diverse \cgls{GW} populations injected into extended data stretches is essential to quantify these impacts, simulating realistic scenarios and avoiding arbitrary signal choices.

In future work, we plan to apply this null-stream method to a large-scale \cgls{MDC} for the triangular \cgls{ET}, simulating realistic \cgls{GW} populations based on established models for \cgls{BBH} mergers, \cgls{BNS} mergers, and other astrophysical sources.
By incorporating \cgls{ET}'s projected sensitivity and generating signals from population distributions, the \cgls{MDC} will enable a comprehensive assessment of the expected calibration error constraints achievable over extended observation periods.
We anticipate quantifying the degree of improvement in relative and absolute calibration precision, particularly at low frequencies (1--1000 Hz).
In this regime, the long-wavelength approximation is sufficiently accurate, ensuring that the sky-independent null stream holds.
The presence of multiple overlapping signals further breaks parameter degeneracies and enhances constraints on calibration errors.
This study will provide key insights into the calibration accuracy attainable for \cgls{ET}, informing network optimization and advancing low-frequency \cgls{GW} science.

\textit{Acknowledgement---} We thank Arthur~Offermans, Justin~Janquart, Peter~T.~H.~Pang, Soumen~Roy, Sumit~Kumar, Thibeau~Wouters and Tito~Dal~Canton for valuable discussion.
M.~W. is supported by the Research Foundation - Flanders (FWO) through Grant No. 11POK24N.
The resources and services used in this work were provided by the VSC (Flemish Supercomputer Center), funded by the Research Foundation - Flanders (FWO) and the Flemish Government.
This work made use of the following software: \texttt{astropy}~\cite{1307.6212, 1801.02634, 2206.14220}, \texttt{matplotlib}~\cite{2007CSE.....9...90H}, \texttt{numpy}~\cite{2006.10256}, \texttt{python}~\cite{python}, \texttt{scipy}~\cite{1907.10121, 2022zndo....595738G}, \texttt{Bilby}~\cite{1811.02042,2024zndo..14025463T}, \texttt{corner.py}~\cite{2016JOSS....1...24F, 2021zndo....591491F}, \texttt{Cython}~\cite{2011CSE....13b..31B}, \texttt{h5py}~\cite{collette_python_hdf5_2014, 2021zndo....594310C}, \texttt{Numba}~\cite{2015llvm.confE...1L, 2022zndo...4343230L}, and \texttt{tqdm}~\cite{2022zndo....595120D}.
This research has made use of the Astrophysics Data System, funded by NASA under Cooperative Agreement 80NSSC21M00561.
This research has made use of \texttt{adstex} (\url{https://github.com/yymao/adstex}).
Software citation information aggregated using \texttt{\href{https://www.tomwagg.com/software-citation-station/}{The Software Citation Station}}~\citep{2406.04405, software-citation-station-zenodo}.

I.~C.~F.~W., F.~C., M.~W., T.~C., H.~N., C.~V.~D.~B., and T.~G.~F.~L. contributed to Conceptualization and Writing -- Review \& Editing.
I.~C.~F.~W. and F.~C. additionally contributed to Data Curation, Formal Analysis, Methodology, Visualization, and Writing -- Original Draft.
I.~C.~F.~W. also contributed to Project Administration, Software, and Validation.
M.~W. also contributed to Methodology and Writing -- Original Draft.
C.~V.~D.~B. and T.~G.~F.~L. also contributed to Funding Acquisition and Supervision.

\textit{Data availability---} The data that support the findings of this article are openly available~\cite{postprocess,data_product}, embargo periods may apply.

\bibliographystyle{apsrev}
\bibliography{references}

\ifprintsupplement
	\clearpage
	\appendix
	\onecolumngrid
	\setcounter{equation}{0}
	\setcounter{figure}{0}
	\setcounter{table}{0}
	\setcounter{section}{0}
	\renewcommand{\theequation}{S\arabic{equation}}
	\renewcommand{\thefigure}{S\arabic{figure}}
	\renewcommand{\thetable}{S\arabic{table}}

	\renewcommand{\theHfigure}{S\arabic{figure}}
	\renewcommand{\theHtable}{S\arabic{table}}
	\renewcommand{\theHequation}{S\arabic{equation}}
	\renewcommand{\theHsection}{S\arabic{section}}

	\section{Supplemental Material}
	\section{Simulation setup}

We simulate a triangular \gls{ET} network to test the Bayesian null-stream method's ability to constrain calibration errors, building on the \gls{BBH} signal setup described in the main text.
Signals are generated using the \texttt{IMRPhenomXPHM} waveform model~\cite{2004.06503}, adjusting luminosity distance to achieve network \glspl{SNR} from 20 to 300 for single-signal cases.
The injection parameters are detailed in Table~\ref{tab:injection_parameters}.
Calibration errors are modeled independently per detector using cubic splines with 10 knots (8--512 Hz, uniform in $\log f$), with knot errors drawn from Gaussian distributions matching LIGO Livingston's calibration uncertainties at GPS time 1256894100~\cite{ligo_virgo_calibration_uncertainty}.
The calibration errors are presented in Fig.~\ref{fig:injection_calibration_error}.
We select the error realization maximizing null stream \gls{SNR}.
For the multi-signal analyses, we simulate cases with one, two, and three distinct \cgls{BBH} signals.
Each signal shares the same intrinsic source parameters (e.g., masses and spins) as detailed in Table~\ref{tab:injection_parameters}, but is assigned a unique sky position and coalescence time, as specified in Table~\ref{tab:injection_parameters_multi_signals}.
The coalescence times are intentionally spaced to prevent temporal overlap, demonstrating that the method's ability to break degeneracies relies on frequency-domain overlap alone.
To ensure a consistent comparison, the luminosity distance of each signal is scaled to contribute equally to a fixed total null-stream \gls{SNR} of 26.3.
This total corresponds to the null-stream \gls{SNR} produced by a single signal with a network \gls{SNR} of 300.
Detector noise follows the uncorrelated ET-D sensitivity curve~\cite{1012.0908}.

An important consideration is the variability of null stream \cgls{SNR} leakage across different calibration error realizations, which can be substantial for LIGO-like calibration uncertainties.
For the \cgls{GW} signal injection with network \gls{SNR} of 300, the distribution of null stream \gls{SNR} across different calibration error realizations spans $11.3^{+5.9}_{-4.0}$ (90\% credible interval).
To clearly demonstrate the method's efficacy in our proof-of-concept study, we deliberately selected the maximum-leakage case.
However, the method remains effective for constraining calibration errors even in more typical scenarios with moderate null stream \cgls{SNR} leakage.

\begin{table}
	\caption{
		Source parameters of the binary black hole signals in the injections.
		The masses are given in the detector frame.}
	\label{tab:injection_parameters}
	\begin{ruledtabular}
		\begin{tabular}{lcc}
			Parameter                    & Value      & Unit        \\
			\hline
			Primary Mass $m_{1}$         & 35.6       & $M_{\odot}$ \\
			Secondary Mass $m_{2}$       & 30.6       & $M_{\odot}$ \\
			Aligned Spin (1) $\chi_{1}$  & 0.3        & --          \\
			Aligned Spin (2) $\chi_{2}$  & 0.36       & --          \\
			Right Ascension $\alpha$     & 1.97       & rad         \\
			Declination $\delta$         & -1.21      & rad         \\
			Polarization Angle $\psi$    & 1.6        & rad         \\
			Inclination Angle $\iota$    & 2.68       & rad         \\
			Coalescence Time $t_{c}$     & 1126259462 & s           \\
			Coalescence Phase $\phi_{c}$ & 0.0        & rad         \\
		\end{tabular}
	\end{ruledtabular}
\end{table}

\begin{table}
	\caption{
		Sky positions and coalescence times of the binary black hole signals in the multi-signal injections.}
	\label{tab:injection_parameters_multi_signals}
	\begin{ruledtabular}
		\begin{tabular}{llll}
			Injection Type & Right Ascension $\alpha$ (rad) & Declination $\delta$ (rad) & Coalescence Time $t_c$ (s)         \\
			\hline
			One-signal     & 1.97                           & -1.21                      & 1126259462                         \\
			Two-signals    & 1.97, 3.89                     & -1.21, -0.37               & 1126259458, 1126259476             \\
			Three-signals  & 1.97, 3.89, 4.77               & -1.21, -0.37, 0.93         & 1126259456, 1126259466, 1126259476 \\
		\end{tabular}
	\end{ruledtabular}
\end{table}

\begin{figure}
	\includegraphics[width=0.6\linewidth]{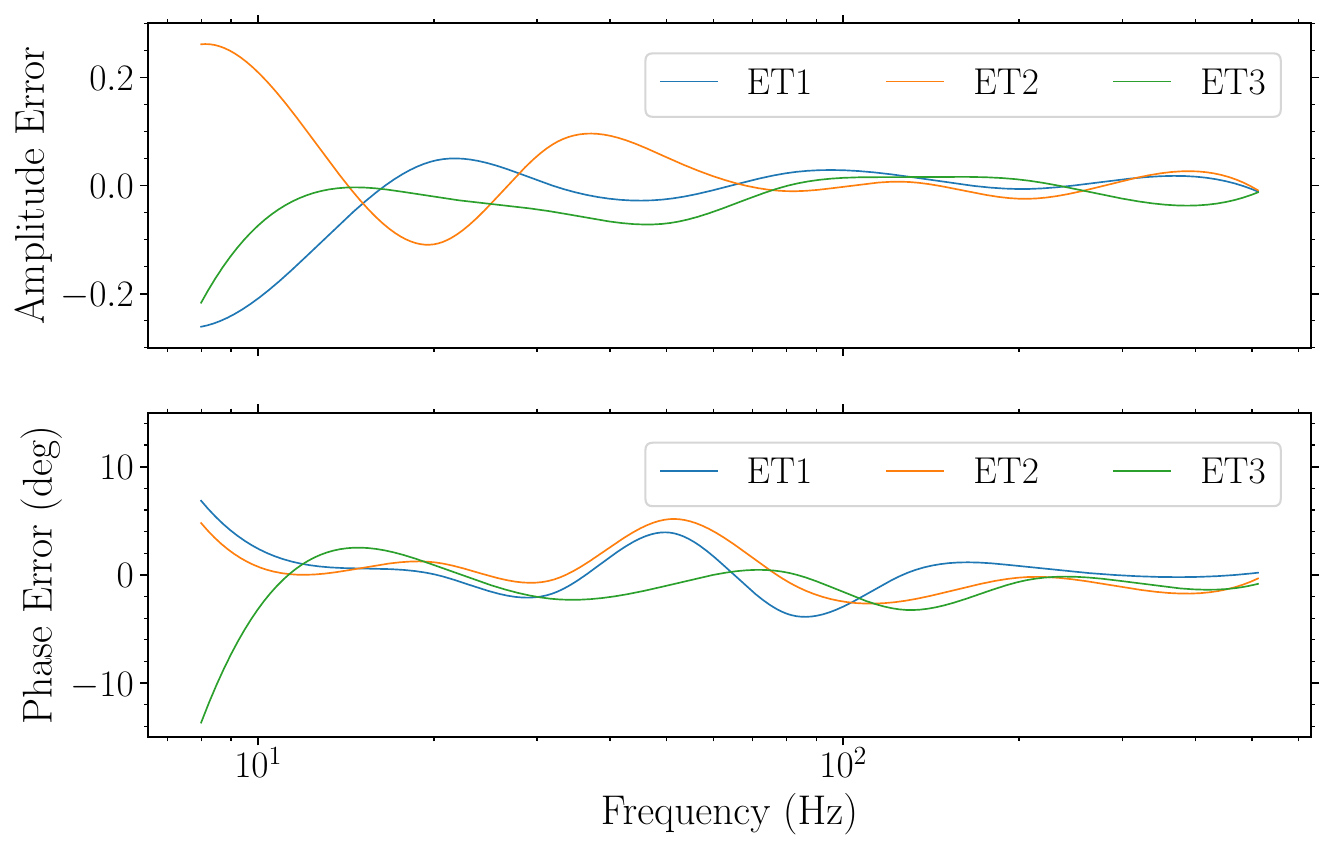}
	\caption{
		The calibration errors applied to the mock data for the three \gls{ET} detectors.
		The top panel shows the amplitude errors, while the bottom panel shows the phase errors.
	}
	\label{fig:injection_calibration_error}
\end{figure}

\section{Degeneracy of the likelihood function}
\label{sec:degeneracy_of_likelihood_function}

The likelihood function of the null stream method exhibits two primary degeneracies that impact calibration error constraints.
First, a common mode degeneracy arises because the likelihood is invariant under a common scaling of calibration errors across detectors~\cite{2009.10212}.
For a closed-geometry network with data
\begin{equation}
	\tilde{\vb{d}}(f) = \vb{C}(f) \sum_i \vb{F}_i \tilde{\vb{h}}_{i}(f) + \tilde{\vb{n}}(f)
\end{equation}
where $\vb{C}(f) = C_{\mathrm{com}}(f)\vb{I}_{3}$ represents the common mode calibration error, $\vb{F}_{i}$ is the antenna pattern matrix for the $i$th source, $\tilde{\vb{h}}_{i}(f)$ are the \gls{GW} polarizations, and $\tilde{\vb{n}}(f)$ is noise, the effect of $C_{\mathrm{com}}(f)$ is equivalent to rescaling the \gls{GW} signal:
\begin{equation}
	\tilde{\vb{d}}(f) = \sum_{i}\vb{F}_{i}
	\left[
		C_{\mathrm{com}}(f)
		\tilde{\vb{h}}_{i}(f)
		\right]
	+
	\tilde{\vb{n}}(f)\,.
\end{equation}
This is evident in the projection matrix.
Define the whitened antenna pattern matrix without calibration errors as
\begin{equation}
	\vb{W}(f) = \sqrt{2\Delta f}\vb{S}_{n}^{-1/2}(f)\vb{F} \,,
\end{equation}
and with common mode errors as
\begin{equation}
	\begin{aligned}
		\vb{W}'(f) & = \sqrt{2\Delta f}\vb{S}_{n}^{-1/2}(f)C_{\mathrm{com}}(f)\vb{I}_{3}\vb{F} \\
		           & = C_{\mathrm{com}}(f)\vb{W}(f) \,.
	\end{aligned}
\end{equation}
The null projection matrix becomes
\begin{equation}
	\begin{aligned}
		 & \vb{P}_{\mathrm{null}}(f) \\
		 & =
		\vb{I}_{3} - \vb{W}'(f)
		\left[\vb{W}'^{\dagger}(f)\vb{W}'(f)\right]^{-1}
		\vb{W}'^{\dagger}(f)         \\
		 & =
		\vb{I}_{3} - \abs{C_{\mathrm{com}}(f)}^{2}
		\vb{W}(f)
		\left[\abs{C_{\mathrm{com}}(f)}^{2}\vb{W}^{\dagger}(f)\vb{W}(f)\right]^{-1}
		\vb{W}^{\dagger}(f)          \\
		 & =
		\vb{I}_{3} - \vb{W}(f)
		\left[\vb{W}^{\dagger}(f)\vb{W}(f)\right]^{-1}
		\vb{W}^{\dagger}(f)          \\
	\end{aligned}
\end{equation}
which is invariant under $C_{\mathrm{com}}(f)$, demonstrating that the likelihood function is unaffected by common mode calibration errors.
Thus, relative calibration errors are better constrained than absolute errors.

The second degeneracy arises for polarized \gls{GW} signals, where $\tilde{h}_{\times}(f) = z\tilde{h}_{+}(f)$ and $z$ is a complex number.
Specifically, the \gls{GW} signals are elliptically polarized where $z = i\epsilon$ when the source is a quasi-circular binary with a low mass ratio, as is the case for most events detected during the first three observing runs of the LIGO and Virgo detectors.
For this scenario, the signal is
\begin{equation}
	\begin{aligned}
		\tilde{\vb{s}}(f)
		 & =
		\vb{C}(f)\vb{F}\tilde{\vb{h}}(f) \\
		 & =
		\vb{C}(f)
		\begin{bmatrix}
			\vb{f}_{+} & \vb{f}_{\times}
		\end{bmatrix}
		\begin{bmatrix}
			\tilde{h}_{+}(f) \\
			\tilde{h}_{\times}(f)
		\end{bmatrix}             \\
		 & =
		\vb{C}(f)
		\begin{bmatrix}
			\vb{f}_{+} & \vb{f}_{\times}
		\end{bmatrix}
		\begin{bmatrix}
			\tilde{h}_{+}(f) \\
			i\epsilon\tilde{h}_{+}(f)
		\end{bmatrix}         \\
		 & =
		\vb{C}(f)
		(\vb{f}_{+} + i\epsilon\vb{f}_{\times})
		\tilde{h}_{+}(f)
	\end{aligned}
\end{equation}
where $\vb{f}_{+}$ and $\vb{f}_{\times}$ are the $3\times 1$ antenna pattern function vectors for the $+$ and $\times$ polarizations, respectively.
This reduces the signal space dimensionality from two to one, yielding two null vectors orthogonal to $\vb{C}(f) (\vb{f}_{+} + i \epsilon \vb{f}_{\times})$, corresponding to two distinct null streams.
This degeneracy broadens the posterior distribution of calibration parameters.
Multiple \gls{GW} signals overlapping in the frequency domain, with differing ellipticities and orientations, break this degeneracy, tightening constraints on calibration errors.

\section{Calibration error constraints at other frequencies}

In the main text, we presented the posterior distributions of relative calibration errors at 32 Hz.
Here, we present the posterior distributions of relative calibration errors across the entire frequency range (8--512 Hz) analyzed, in Fig.~\ref{fig:calibration_envelope_comparison_3_signals}, for the three-signal case with a null stream \gls{SNR} of 26.3.
The shaded regions represent 90\% credible intervals, and the dashed lines indicate the true values.
Across the frequency range, the posteriors are significantly tighter than the priors, accurately recovering the true values, demonstrating the method's effectiveness in constraining calibration errors throughout the analyzed frequency band.
Remarkably, at low frequencies around 10 Hz, the true relative calibration errors for the amplitude between ET2 and ET1 are outside the 90\% credible interval of the prior, yet the posterior successfully recovers the true value, showcasing the method's capability to enhance calibration precision in the situation where the state of the interferometers at the moment deviates from the instrumental measurements over a longer period.
The deviation of the injected calibration errors from the 90\% credible interval of the prior at certain frequencies is due to the selection of the calibration error realization that maximizes the null stream \gls{SNR}.
This demonstrates the method's ability to improve calibration precision even when the actual calibration errors deviate significantly from prior expectations.

\begin{figure}
	\centering
	\includegraphics[width=0.497\linewidth]{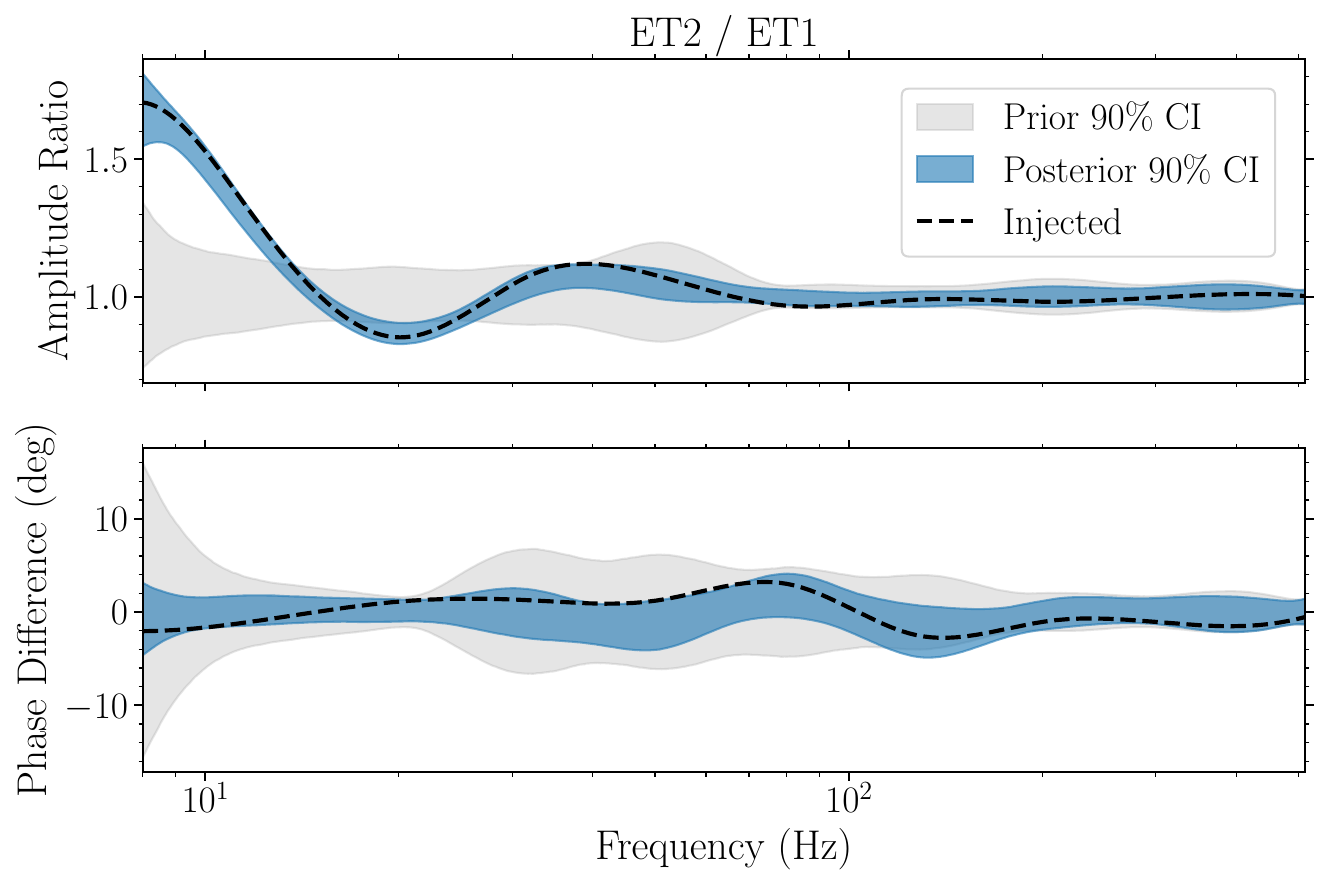}
	\includegraphics[width=0.497\linewidth]{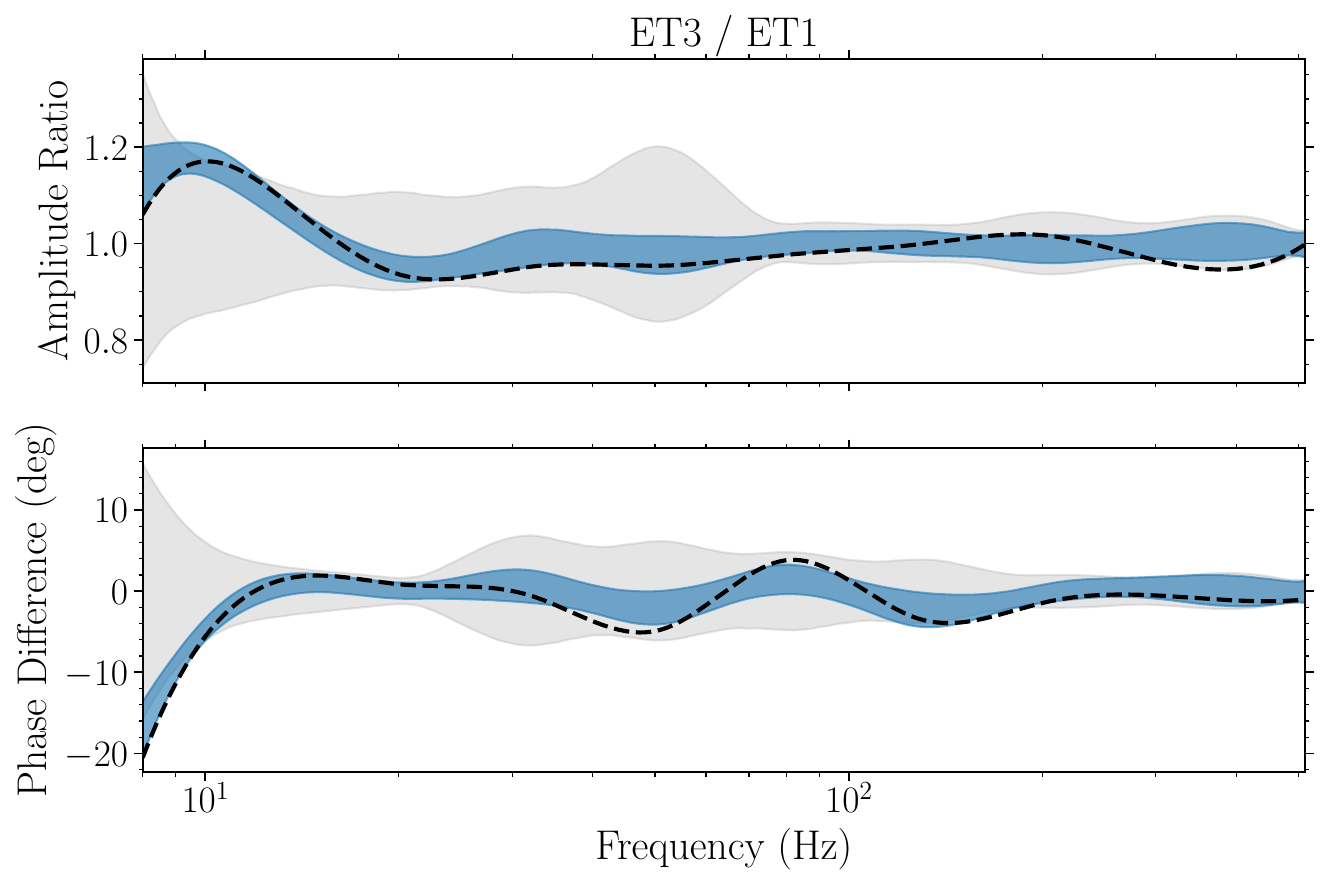}
	\caption{
		The plot shows the posterior distributions for the amplitude ratio $(1 + \delta A_{j}) / (1 + \delta A_{1})$ and phase difference $\delta \phi_{j} - \delta \phi_{1}$ of calibration errors for the $j$th \gls{ET} detector compared to the first, across the frequency range 8--512 Hz, where $\delta A_{j}$ and $\delta \phi_{j}$ are the amplitude and phase errors, respectively.
		The left panel compares the second detector (ET2) to the first (ET1), and the right panel compares the third (ET3) to the first (ET1).
		The shaded regions represent 90\% credible intervals.
		The results are from the three-signal case with a null stream \gls{SNR} of 26.3.
		Across the frequency range, the posteriors are significantly tighter than the priors, accurately recovering the true values (dashed lines).
	}
	\label{fig:calibration_envelope_comparison_3_signals}
\end{figure}

\fi

\end{document}